\documentclass[aps,prl,floats,floatfix,twocolumn]{revtex4-1}

\usepackage{color} %\textcolor{red} \color[RGB]{102,0,0}
\usepackage{latexsym}
\usepackage{amsmath} 
\usepackage{amssymb} 
\usepackage{bm}
\usepackage{wasysym}
\usepackage[normalem]{ulem}
\usepackage{graphicx}
\usepackage{epstopdf}
\graphicspath{{./}{./Figs/}}
\usepackage[percent]{overpic}
\usepackage[colorlinks,linkcolor=blue,urlcolor=blue,bookmarks=true,pdftitle={}]{hyperref}

\newcommand{\hide}[1]{}

\newcommand{\Eq}[1]  {{\textcolor{blue}{Eq.}}~(\ref{#1})}

\makeatother

\begin{document}

\title{Anyonic-parity-time symmetry in complex-coupled lasers}

\author{Geva Arwas$^{*}$, Sagie Gadasi$^{*}$, Igor Gershenzon, Asher Friesem, Nir Davidson, Oren Raz}

\affiliation{
Department of Physics of Complex Systems, Weizmann Institute of Science, Rehovot 7610001, Israel
}

\begin{abstract}

Non-Hermitian Hamiltonians, and particularly parity-time (PT) and anti-PT symmetric Hamiltonians, play an important role in many branches of physics, from quantum mechanics to optical systems and acoustics. Both the PT and anti-PT symmetries are specific instances of a broader class known as anyonic-PT symmetry, where the Hamiltonian and the PT operator satisfy a generalized commutation relation. Here, we study theoretically these novel symmetries and demonstrate them experimentally in coupled lasers systems. We resort to complex coupling of mixed dispersive and dissipative nature, which allows unprecedented control on the location in parameter space where the symmetry and symmetry-breaking occur. Moreover, tuning the coupling in the same physical system, allows us to realize the special cases of PT and anti-PT symmetries. In a more general perspective, we present and experimentally validate a new relation between laser synchronization and the symmetry of the underlying non-Hermitian Hamiltonian.
\end{abstract}

\maketitle

\date{}

\section*{Introduction}

Hamiltonian quantum theory provides an excellent description of isolated and closed systems. The Hermiticity of the Hamiltonian operator assures that the probability flows only between the various states of the system.
To model open systems, which can exchange probability and heat with their environment,  non-Hermitian effective Hamiltonians are commonly used \cite{nimrod1,bender2007making}.
These non-Hermitian Hamiltonians can display complex spectra and non-orthogonal eigenmodes, leading to several unusual properties. A notable example is the spectral degeneracy known as exceptional point (EP)  \cite{berry2004physics,heiss2012physics}, where two or more eigenmodes coalesce.
Additional important examples are several novel symmetries that can appear only in non-Hermitian Hamiltonians. In particular, parity-time (PT) symmetric systems \cite{bender1998real} have gained much interest, due to their unique and counter-intuitive properties. Importantly, these Hamiltonians can have a real valued spectrum even when non-Hermitian. This property, which is often referred to as `pseudo-Hermiticity' \cite{mostafazadeh2002pseudo}, is not limited to PT-symmetric systems, and plays an important role in complex extensions of quantum theory \cite{bender2002complex,rotter2009non}.

As many concepts that originated from non-Hermitian physics, the PT symmetry was first developed in the context of quantum mechanics, but turned out to be a powerful tool in several other branches of physics, including electronic \cite{PhysRevA.84.040101}, acoustic \cite{zhu2014p,fleury2015invisible},
and in particular -- optics and photonics \cite{feng2017non,ozdemir2019parity,miri2019exceptional,klaiman2008visualization}. 
In optical systems, the parity operator spatially reflects the system and the `time reversal' operation interchanges the gain with the loss.
The PT symmetry then emerges from a gain-loss balance \cite{ruter2010observation,guo2009observation,el2007theory,chong2011p}, which can keep the  Hamiltonian invariant under the combined operation.
This offers a novel control over the light's spatial profile and the device's transmission properties \cite{lin2011unidirectional,makris2008beam,longhi2010pt}.
A PT symmetric system can be in two different phases, with a sharp symmetry-breaking \cite{guo2009observation,peng2014parity} transition between the two.
In the vicinity of the symmetry-breaking, the system shows extreme sensitivity to small perturbations, a highly promising feature for sensing applications \cite{wiersig2014enhancing,chen2017exceptional,hokmabadi2019non}.
PT symmetric lasers have been extensively studied, where the symmetry breaking was utilized to generate a robust single-mode operation \cite{feng2014single,hodaei2014parity}, reversing the pump dependence \cite{brandstetter2014reversing} or loss induced lasing \cite{peng2014loss}.
A closely related concept which was recently demonstrated in \cite{anti1,anti2,anti3} is anti-PT symmetry. Optical anti-PT systems offer additional methods to control light which can be used, for example, to generate a refractionless propagation.

In this paper, we generalize the PT and anti-PT symmetries into an anyonic-PT.
The general anyonic-PT, and the two special cases of PT and anti-PT symmetries, are all experimentally realized in a single physical system of two coupled lasers in a degenerate cavity. The specific type of anyonic-PT symmetry is tuned by controlling the phase of the complex coupling between the lasers \cite{cao2019complex}. 
We show both experimentally and theoretically how the symmetry is manifested, and broken, along a line in the lasers' relative loss and frequency parameter space. 
Furthermore, by taking into account the nonlinearity of the laser system, we relate our results to the physics of synchronization, and show how the various non-Hermitian symmetries are manifested. 

Anyonic-PT symmetric Hamiltonians satisfy
\begin{eqnarray}
\mathcal{PTH} \ = \ \mathrm{e}^{-2 i \beta} \mathcal{HPT}  \label{e1}
\end{eqnarray}
where $\mathcal{P}$ and $\mathcal{T}$ are the parity and time-reversal operators, and $\beta$ is a real constant. The specific cases $\beta=0$ and $\beta=\pi$ correspond to the standard PT-symmetry,
where the Hamiltonian satisfy $[ \mathcal{PT},\mathcal{H}]=0$.
The cases $\beta=\pm \pi/2$  correspond to the anti-PT symmetry, where the Hamiltonian anti-commutes with the PT operation, $ \{ \mathcal{PT},\mathcal{H} \}=0$.
The (anti-)PT commutation relations resemble the famous (fermionic) bosonic commutation relations. In this spirit, the general case was recently named anyonic-PT symmetry \cite{anyonic1,anyonic2}. 
It should be clear that this terminology does not suggest the existence of anyonic quasi-particles in the system, but rather a formal analogy with the commutation relation \Eq{e1}.

The anyonic-PT symmetry is a novel way to manipulate light. For example, we demonstrate how it can be used to control the phases and the intensities of two coupled lasers, which can have arbitrary losses or frequencies. Furthermore, our methods enable to manipulate the location of the EPs in parameter space. In this way, the system can be tuned to the vicinity of an EP, without having to modify the frequencies or the losses, as required in the case of PT symmetric systems.

Non-Hermitian physics also suggests a new perspective on synchronization. A common wisdom \cite{synchronizationbook} states that a stable unique synchronized state for two phase oscillators with different frequencies can be obtained only through a dissipative coupling mechanism \cite{ding2019dispersive,avi}, and that their frequency detuning must not exceed the strength of this coupling. Contrary to the expectation, we show how by controlling the lasers’ loss, synchronization is possible for larger detuning, and even when the coupling is purely dispersive. The allowed frequency range follows a universal relation which we validate experimentally. We show that for all coupling types the de-synchronization generally results from a pseudo-Hermiticity symmetry.

\section*{Results}

Although our analysis can be straightforwardly generalized to many other systems, here we focus on the specific experimental system used in what follows. It is composed of two coupled lasers with complex electric fields $E_1$ and $E_2$, relative loss $\Delta \alpha =  (\alpha_1 - \alpha_2 )/2$ and frequency detuning  $\Delta \Omega =  (\Omega_1 - \Omega_2)/2 $ (see Fig. 1). 
In each cavity round-trip, part of the light from each laser is coupled, in a symmetric manner, into the other laser (see Fig. 2 and Methods, experimental setup section). 
 The evolution of the coupled laser fields is given by the laser rate equations \cite{rogister2004power} (see Methods, laser rate equations section), that we write in a vectorial form 
\begin{eqnarray}
i \frac{d \mathbf{E}}{dt}  \  = \ [ i G(\mathbf{E} ,t) + \Omega_0 - i \alpha_0 + \mathcal{H} ] \mathbf{E} ,
\label{e6}
\end{eqnarray}
where $\mathbf{E} = [E_1,E_2]^T$, $ G(\mathbf{E} ,t) $ is a diagonal matrix which represents the nonlinaer gain, $\alpha_0 = (\alpha_1 + \alpha_2 )/2$ and $\Omega_0 =  (\Omega_1 + \Omega_2)/2 $ are scalar constants representing the lasers' average loss and frequency, correspondingly. $\mathcal{H}$ is an effective traceless Hamiltonian matrix, parametrized by
\begin{eqnarray}
\mathcal{H} \ = \ 
\left( 
\begin{array}{cccc}
 z   & \kappa \mathrm{e}^{i \beta } \\
 \kappa \mathrm{e}^{i \beta } & -  z   \\
\end{array}
\right) , \label{e2}
\end{eqnarray}
where ${z= \Delta \Omega - i \Delta \alpha }$.
The frequency and the loss of each laser (relative to their average values), are the real and imaginary parts of the Hamiltonian’s diagonal, while the off-diagonal terms describe the complex coupling  between the lasers, obtained by calculating a spatial overlap between the two laser fields (see Methods). Since the coupling in our system is symmetric, the two coefficients are identical.
The magnitude of the coupling is given by $\kappa$ and its phase by $\beta$. 
%As discussed below, the angle $\beta$ in the above definition of the Hamiltonian is identical to the $\beta$ in \Eq{e1}. 
%
This type of coupling, unless purely real, breaks Hermiticity and generically leads to a non-conservative dynamic, even in the absence of gain or loss in the system. We therefore refer to real coupling ($\beta=0,\pi$) as dispersive, and to imaginary coupling ($\beta=\pm \pi/2$) as dissipative. For complex coupling the ratio between the dispersive and dissipative parts is set by $\beta$.

For simplicity, we begin the analysis with an approximated linear description of the dynamics, where 
we replace $ G(\mathbf{E} ,t) $ by a constant value $ G $, and therefore neglect the spatial and temporal dependence of the gain medium.
Under this approximation, the steady-state ($d \mathbf{E}/{dt}=0$) solutions of \Eq{e6} are given by the eigenmodes of $\mathcal{H}$. We address the full non-linear dynamics in a later section.

\begin{figure*}[t]
\includegraphics[width=0.99\hsize]{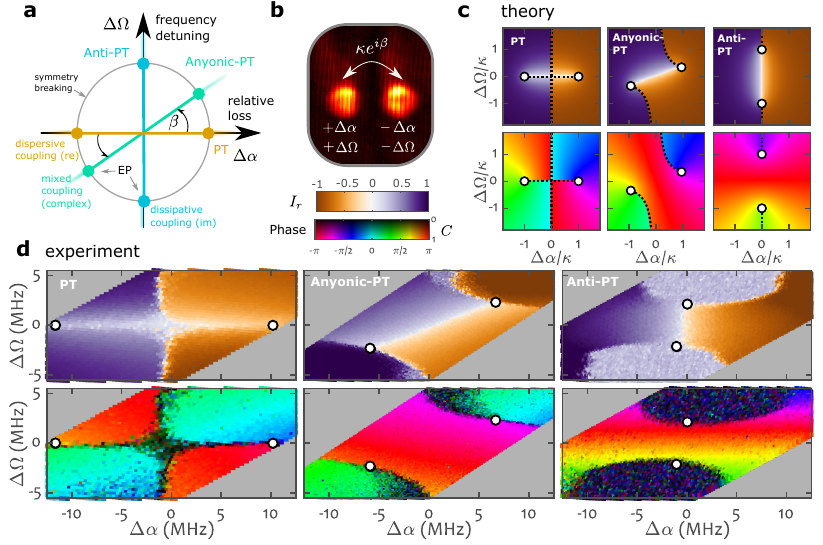}
\caption{
\textbf{Non-Hermitian symmetries and complex coupling.} (a) Schematic diagram of the anyonic-PT symmetry line. The coupling type – from dispersive to dissipative – given by $\beta$, rotates the symmetry line in the $(\Delta \alpha , \Delta \Omega)$ plane. 
The dots represents the location of the EPs, on a circle of radius $\kappa$, where the symmetry breaks in each case.
(b) The experimentally controllable parameters. The background show a single shot of the two coupled lasers. The interference fringes are used to extract the relative phase in each measurement.
(c) Theoretical plots of the relative intensity $I_{r}$ (top row) and phase difference of the lasers $\phi$ (bottom row), as given by the $V_+$ eigenvector of the effective Hamiltonian. In each panel the coupling is fixed and a $(\Delta \alpha , \Delta \Omega)$ regime diagram is plotted. The different columns are for $\beta = \pi, -0.88 \pi , - \pi/2 $, corresponding to purely dispersive, complex, and purely dissipative couplings, respectively. The dots mark the location of the EP. The dashed lines represents the condition for pseudo-Hermiticity of the Hamiltonian. (d) Experimental measurements of the relative intensity $2I_{r}$ (top row) and phase difference of the lasers $\phi$ (bottom row). The different columns correspond to $\kappa \approx 10.9,6.7,2.2 $ (MHz) respectively, with approximately the same $\beta$ values of (c). The phase difference values are shaded in accordance to their measured phase coherence $ C= | \langle \mathrm{e}^{ i \phi}  \rangle |  $, where black regions correspond to poor coherence and lack of synchronization. Gray areas are experimentally inaccessible. }
\end{figure*}

\begin{figure*}
\includegraphics[width=0.98\hsize]{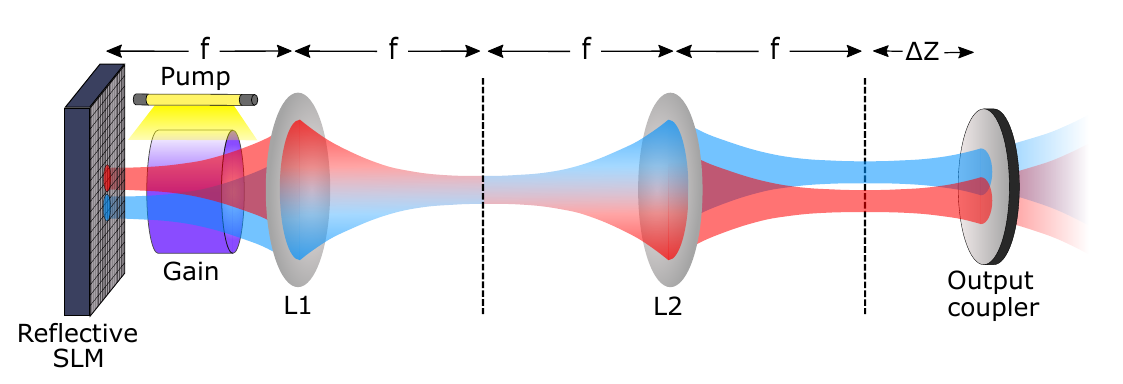}
\caption{ \textbf{Degenerate cavity laser arrangement.} It is comprised of two lenses (L1 and L2) that form a 4f telescope configuration, reflective SLM for forming two lasers, an output-coupler, and a Nd:YAG gain medium that is pumped by a Xenon flash lamp. Coupling between the two lasers is achieved by displacing the output-coupler a distance $\Delta z$. The SLM is used to apply frequency detuning and relative loss between the lasers. }
\end{figure*}

The Hamiltonian in \Eq{e2} is anyonic-PT symmetric, i.e., it satisfies \Eq{e1}, provided that (see Methods):
\begin{eqnarray}
 \tan \beta \ = \ \frac{\Delta \Omega}{\Delta\alpha} .
 \label{e3}
\end{eqnarray}
For a fixed $\beta$, this corresponds to a tilted straight line in the $(\Delta \alpha , \Delta \Omega)$ parameter space, as schematically shown in Fig. 1(a). We refer to this line as the anyonic-PT symmetry line. To show how the symmetry is manifested in the spectrum of the system, the eigenvalues and (non-normalized) eigenmodes of $\mathcal{H}$ are written as:
\begin{eqnarray}
\lambda_\pm \ = \ \pm \ i \sqrt{-\kappa^2 \mathrm{e}^{2 i  \beta } -z^2 }   , \ 
V_\pm \ = \ \left[ \ \frac{z + \lambda_\pm }{  \kappa \mathrm{e}^{i \beta }  } , 1 \right]^T. \label{e4}
\end{eqnarray}
The real and imaginary parts of $\lambda_\pm$ correspond (up to a constant) to the frequency and the loss of the eigenmodes. $\lambda_+$ and $V_+$  represent the less lossy eigenmode.
In Fig. 1(c), we plot theoretical regime diagrams for the relative phase $\phi = \phi_1 - \phi_2$ and intensity $I_r = (A_1^2-A_2^2)/(A_1^2+A_2^2) $ of the two laser fields $E_i= A_i \mathrm{e}^{  i \phi_i}$, when the system is in the $V_+$ mode. % in \Eq{e4}.
In each image $\beta$ is fixed, and a $(\Delta \alpha , \Delta \Omega)$ regime diagram (both in units of the coupling magnitude $\kappa$) is plotted. The different values of $\beta$ correspond to a dispersive, complex, or dissipative coupling.

First, we consider the symmetry line that is clearly visible for all values of $\beta$ (see Fig. 1(a)). In the dispersive case ( $\beta=\pi$ , Fig. 1(c) left) the well known  PT-symmetry line \cite{ozdemir2019parity} is located at $\Delta \Omega=0$. Along this line, in the so called `unbroken' or `exact' regime $|\Delta \alpha | < \kappa $, the two lasers maintain equal intensity, while only their relative phase is changed. In the `broken' regime $|\Delta \alpha | > \kappa $, one laser becomes stronger, while their relative phase is fixed at $\pm \pi/2$.
Here, the Hamiltonian and the $\mathcal{PT}$ operator do not share the same eigenmodes, even though the operators commute. At the symmetry breaking points, $|\Delta \alpha |=\pm \kappa$ the eigenvalues $\lambda_\pm$, as well as the eigenmodes $V_\pm$, coalesce. This type of non-Hermitian degeneracy is known as the exceptional-point (EP).

In the dissipative case, ($\beta=-\pi/2$, Fig. 1(c), right), the anti-PT symmetry line is located at $\Delta \alpha=0$, where $|\Delta \Omega | < \kappa $ and ${|\Delta \Omega | > \kappa }$ correspond to the `unbroken' and `broken' phases. Here the EPs are located at $ \Delta \Omega  = \pm \kappa $.

The new anyonic-PT symmetry is presented in the middle panel of Fig. 1(c). 
In this case the coupling is complex valued with $\beta=-0.88 \pi$, and the anyonic-PT symmetry line is $\beta$-rotated.
As in the PT-symmetric case, the two lasers have equal amplitudes along the anyonic-PT symmetry line
in the `unbroken' phase, and the symmetry breaking occurs at the EPs,
located at ${(\Delta \alpha , \Delta \Omega)=\pm (\kappa \cos \beta , \kappa \sin \beta)}$.

Beyond the symmetry lines, all phase diagrams reveal a rich structure. 
In particular, they show a discontinuity jump in the phase and/or intensity of the lasers across the dashed lines, which we address later on. Although the anyonic-PT symmetry line is tilted by an angle $\beta$ from the PT symmetric case, 
the full regime diagram is not rotated in a trivial way, manifesting the different physical roles of $\Delta \alpha$ and $\Delta \Omega$.
When the coupling is complex, there is no symmetry associated with reflections around the $\Delta \alpha=0$ axis. Strikingly, $\Delta \alpha >0$ does not always result in $I_r>0$, i.e, the more lossy laser might have a stronger intensity.

The experimental measurements are performed for two coupled lasers in a degenerate cavity.
To control $\beta$, we adjust the distance between the center of the two laser spots, (see Fig. 1(b) and Methods). Note that by doing so, we also change $\kappa$ as the magnitude and phase of the coupling are not independent. Once $\beta$ is fixed, we use an intra-cavity digital mirror to apply a phase and amplitude mask which allows us to control the loss and the frequency of each laser individually and to scan $\Delta \alpha$ and $\Delta \Omega$ values. Finally, we measure the intensity of the two lasers and their relative phase by interfering the two beams (see Methods). 

In Fig. 1(d) we show the experimental results of the relative phase and intensity of the two coupled lasers.
The symmetry line can be easily identified in each case by its unbroken phase of equal intensities, as well as the sharp symmetry breaking–a signature of an EP.
We find a remarkable agreement with the theoretical predictions of the $V_+$ mode in Fig. 1(c).
As opposed to a coherently driven system, where different eigenmodes can be resonantly excited,
for our coupled lasers, the lasing state is the winner of a mode competition \cite{siegman1986lasers}, well approximated by  the lower loss eigenmode  \cite{phaseretrieval,house}.

Yet, the laser system is inherently nonlinear. The diagrams in Fig. 1(d) contain regions of low-coherence, where the black color indicates that the relative phase between the lasers is not well defined. In these regions the lasers do not synchronize due to the coexistence of degenerate modes, as we confirm by nonlinear simulations (see Supplementary).
The lack of phase coherence can not be captured by a linear analysis, but, as we show below, results from a pseudo-Hermiticity symmetry of $\mathcal{H}$ (along the dashed lines in Fig. 1(c)).
We note that at the vicinity of the EPs, linewidth broadening \cite{linewidth1,linewidth2} can also affect the phase coherence of the lasers \cite{linewidth3}.

We now explore the behaviour along the anyonic-PT symmetry line in more details. Fig. 3(a) shows the amplitude ratio along the line for various values of $\beta$, as a function of the frequency detuning.
The unbroken phase is bounded by $\Delta \Omega = \pm \kappa \sin \beta $, which also mark the location of the two EPs, with a sharp symmetry breaking.
A clear plateau of equal amplitudes in the unbroken phase is shown, in perfect agreement with the theory.
In fact, as long as the anyonic-PT symmetry is not broken, the linear modes are also exact solutions of the full non-linear laser rate equations, as explained below. The different place of symmetry breaking for each $\beta$ demonstrates our ability to control the location of the EPs in the $(\Delta \alpha , \Delta \Omega)$ parameter space. Operating at the vicinity of the EPs, enhances the sensitivity of the system to perturbations, which may be advantageous for sensing applications \cite{wiersig2014enhancing,chen2017exceptional,hokmabadi2019non}.

Along the symmetry line, from \Eq{e4} it follows that
\begin{eqnarray}
 \Delta \Omega \ = \  - \kappa \sin \beta \sin \phi , \label{e5}
\end{eqnarray}
such that $\phi$ varies in the unbroken phase, between the two EPs, from $\pi/2$ to $-\pi/2$ and then remains constant in the broken phase, beyond the EPs.
In Fig. 3(b) we plot the measured phases as a function of $\Delta \Omega$, scaled by a factor of $ \kappa \sin \beta $. 
The measured $\phi$, plotted here for different $\beta$, show a similar trend upon this scaling,
in agreement with \Eq{e5}. 
The large errorbars at the vicinity of the symmetry breaking indicates the dramatic enhancement of the noise at the EPs.

\begin{figure}
\includegraphics[width=0.94\hsize]{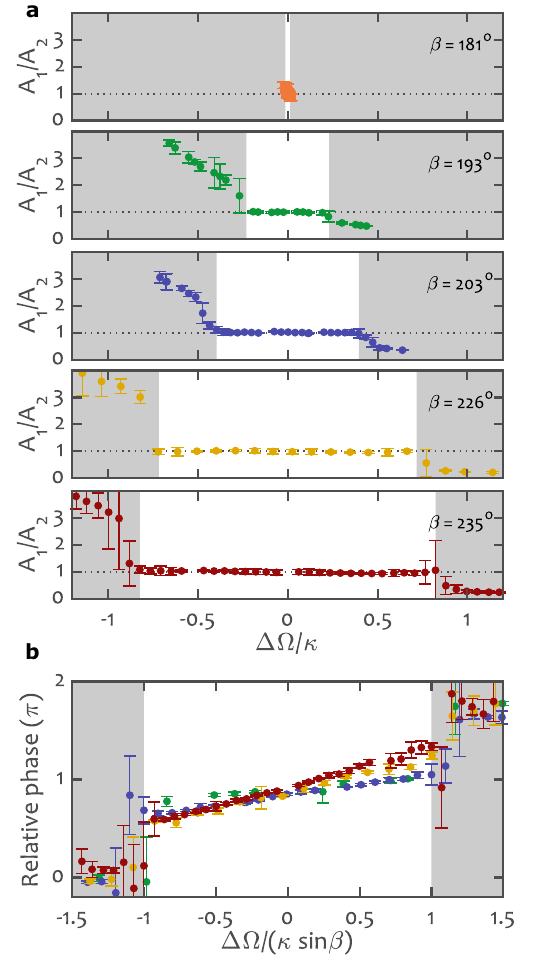}
\caption{ \textbf{The anyonic-PT symmetry line.} (a) Experimental measurements of the amplitude ratio of the lasers $A_1/A_2$ along the anyonic-PT symmetry line, as a function of the detunning $\Delta \Omega$. Each panel corresponds to different value of $\beta$, as indicated in the figure. The anyonic-PT unbroken regime is identified by the equal amplitude plateau, while in the anyonic-PT broken regime (shaded) one laser has a larger amplitude. The EPs, located at $\Delta \Omega = \pm \kappa \sin \beta  $ mark the place of symmetry breaking for each $\beta$. (b) Experimental measurements of the relative phase of the two lasers along the symmetry line. Here the detuning is given in units of $\kappa \sin \beta $. Upon this scaling, the location of the EPs and the onset of the symmetry breaking is the same for all $\beta$. The different colors correspond to the same $\beta$ values of (a). }
\end{figure}

We now turn to address the general structure of the regime diagrams. The low-coherence in the relative phase shown in Fig. 1(d) indicates that the lasers fail to synchronize. We next show that this is a manifestation of \emph{pseudo-Hermiticity}. This notion refers to operators that are non-Hermitian, but nevertheless have a pure real spectrum \cite{mostafazadeh2002pseudo}.
Here, the condition for $\mathcal{H}$ to be pseudo-Hermitian, i.e., $ Im [ \lambda_- ] =  Im [ \lambda_+ ]=0 $, is shown by dashed lines in  Fig. 1(c), given by
\begin{eqnarray}
\Delta \alpha \Delta \Omega = \frac{\kappa^2 \sin 2 \beta }{2 } , \label{epseudo}
\end{eqnarray}
with $  \kappa^2 \cos 2 \beta - \Delta \alpha^2  + \Delta \Omega^2 > 0 $. For the purely dispersive and purely dissipative cases, this becomes a trivial straight lines, overlapping with the unbroken-PT and the broken-anti-PT phases, where the eigenvalues are real. The complex coupling is different in this regard: here \Eq{epseudo} does not overlap with the anyonic-PT symmetry line, where the spectrum is always complex. 

For all coupling types, we see a sharp discontinuity jump in the phase and/or intensity in the theoretical diagrams of Fig. 1(c) across the pseudo-Hermiticity lines.
Mathematically, the discontinuity can be attributed to a branch cut of the complex $\lambda_\pm$ and $V_\pm$ functions.
In the experimental results in Fig. 1(d), the nonlinearity and the noise in the system broaden this discontinuity, and generate low coherence regions around the pseudo-hermiticity lines.
Pseudo-Hermiticity implies that all modes have the same loss resulting in poor phase coherence, due to the coexistence of different lasing modes. 
For  purely dispersive  coupling Fig. 1(d, left) this results in a unique `$+$' shape. Here, in addition to the pseudo-Hermiticity in the horizontal unbroken-PT symmetry line, along the vertical $\Delta \alpha=0$ line, the Hamiltonian is truly Hermitian, trivially having real eigenvalues.
This `$+$' shape is extremely fragile – a slight change of $\beta$ from $\pi$ changes the regime diagram dramatically (see Supplementary).

So far we showed how the breakdown of synchronization is linked to the symmetry of the underlying effective linear Hamiltonian. 
We now extend the analysis to include the nonlinear effects in the system. With these, we can quantify the conditions for synchronization and understand how the nonlinearity affects the non-Hermitian symmetries of the Hamiltonian.

The nonlinearity in the coupled laser system results from the term $G(\mathbf{E} ,t)$ in \Eq{e6}, which describes the dynamics of the gain medium.
In a steady state, if one exists, the gain of each laser takes a constant value of $G_i  = P/(1+ | E_i |^2)  $, where $P$ is the pump strength and $|E_i|^2$ is the laser intensity, normalized to its saturation value (see Methods). 
In the anyonic-PT unbroken phase, both lasers have the same amplitude, yielding $G_1=G_2$. Hence, the solutions of \Eq{e6} are exact eigenmodes of $\mathcal{H}$  \cite{ge2016nonlinear,teimourpour2017robustness} and the anyonic-PT unbroken phase is not affected by the non-linearity.
The pseudo-Hermiticity symmetry line, however, is not robust in this sense, and is expected to vary with different strengths of nonlinearity in the system.

In the broken symmetry phase, or away from the symmetry line, the lasers do not necessary synchronize, or even lase at all.
The condition for a synchronized steady state is that the system converges to a stable fixed point of \Eq{e6}, namely that the two laser fields have fixed intensities and oscillate in synchrony, with a `locked' relative phase. The dynamic of the phase difference between the lasers is given by (see Methods):
\begin{eqnarray}
\frac{d\phi}{dt} = -2 \Delta \Omega +  \kappa \left[ \frac{A_1}{A_2} \cos(\phi + \beta ) - \frac{A_2}{A_1} \cos(\phi - \beta ) \right] \label{e7}
\end{eqnarray}
Note that this relation only depends on the amplitude ratio. The loss $\Delta \alpha$ and the gain $G(\mathbf{E} ,t)$ do not appear here. Therefore it also applies to many different coupled oscillatory systems and other types of nonlinearities.
In Fig. 4 we show the   $(A_1/A_2, \Delta \Omega)$ regime diagram  of the measured synchronization for the same values of $\beta$ as in Fig. 1.

\begin{figure}
\includegraphics[width=0.98\hsize]{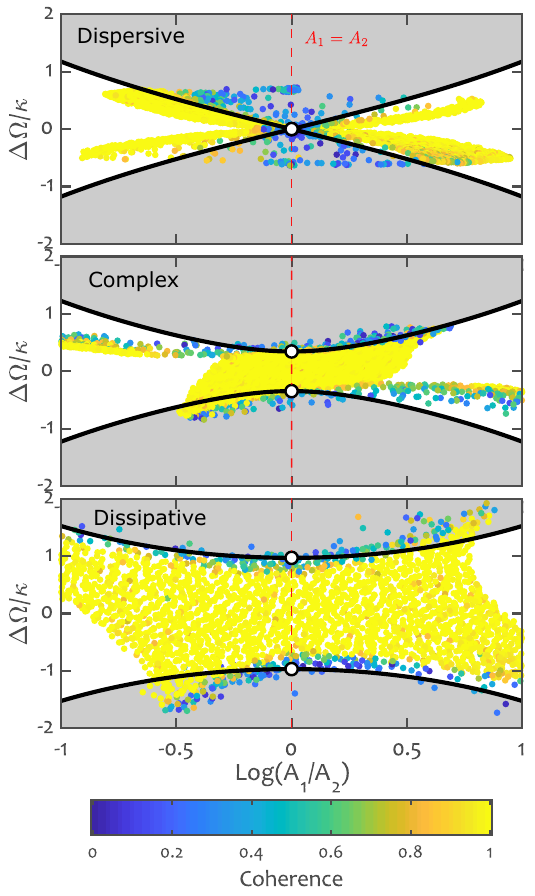}
\caption{\textbf{Synchronization regions.} The different panels correspond to $\beta = \pi, -0.88 \pi , - \pi/2 $, as in Fig. 1(d). The black lines are the theoretical synchronization criteria of \Eq{e9}, while the scattered points show the experimental data. The color represents the measured phase coherence, where yellow color indicates perfect synchronization. Each data point is the average of 10 measurements. The white points show the $\Delta \Omega$ value of the EPs, which border the frequency range for uniform amplitude synchronization.}
\end{figure}

Let us first discuss synchronization with equal amplitudes $A_1=A_2$ (red line in Fig. 4). Here, the condition $d\phi/ dt=0$ in \Eq{e7}  reduces to \Eq{e5} and  coincides with the phase along the anyonic-PT symmetry line. Hence, equal amplitude synchronization is possible only if $|\Delta \Omega |$ is smaller than $\kappa \sin \beta $, the dissipative part of the coupling. Therefore, the anyonic-PT symmetry must be unbroken.
For purely dissipative coupling, this reduces to the well know condition  for synchronization $ |\Delta \Omega | < \kappa  $  \cite{synchronizationbook}, and the border coincides with the anti-PT symmetric EPs \cite{ding2019dispersive}.
Conversely, when the dissipative part vanishes, as in the PT-symmetric case, uniform amplitude synchronization is impossible.

In the case of unequal amplitudes, \Eq{e7} implies that synchronization is possible, in principle, for every $\Delta \Omega$ and $\beta$. The synchronization condition in the region is (see Methods):
\begin{eqnarray}
\left|  \Delta \Omega \right| 
\  \leq \  \frac{\kappa}{2}
\left[ \left( \frac{A_1}{A_2} \right)^2 + \ \left( \frac{A_2}{A_1} \right)^2 - \ 2 \cos{2 \beta} \right]^{1/2} , \label{e9}
\end{eqnarray}
indicated by the black lines in Fig. 4.
For all values of $\beta$, we see a good agreement between the regions of high coherence and the synchronization criteria of \Eq{e9}.
In particular, larger  frequency detunings, require larger amplitude ratio for the lasers to synchronize.
For purely dispersive coupling, we get a unique linear behavior at the origin.
Paradoxically, here it is more difficult to synchronize the lasers when they have the same frequency. This is because for dispersive coupling with $\Delta \Omega=0$, the PT-symmetry, until broken, forces equal amplitudes, preventing the lasers from synchronizing.

\section*{Discussion}

To conclude, we demonstrated and investigated anyonic-PT symmetries using a degenerate cavity laser.
We provided detailed regime diagrams for the case of two coupled lasers, showing how the novel anyonic-PT symmetry, as well as the special cases of PT and anti-PT symmetries, are manifested.
At the heart of our experiments is the ability to control the nature of the coupling – from purely dispersive to purely dissipative. 
The mixed case, where the coupling is complex, presents a rich structure that can be advantageous for future applications.
The presence of a robust symmetry line in parameter space, which depends on both the frequency and the relative loss between the lasers, can be used for calibration, sensing applications, or to detect a frequency-loss correlated noise.
Furthermore, by controlling the coupling, one can control the location of the EPs in parameter space.

An interesting direction for future research is the possibility to dynamically alter the coupling type. For instance, by dynamically changing the coupling from purely dispersive to purely dissipative, A PT symmetric system can be transformed into an anti-PT symmetric system, continuously. In our degenerate laser cavity, this could be achieved by dynamically changing the distance between the lasers.
Furthermore, by dynamically changing the coupling type, the EP itself can move along a circle around a fixed point in parameter space, rather than encircling the EP which is of recent interest \cite{doppler2016dynamically}. 

While synchronization is inherently a nonlinear phenomenon, we demonstrated how the non-Hermitian (linear) framework can provide valuable insight. The lasers fail to synchronize due to the reality of the spectrum's pseudo-Hermiticity symmetry –- as all modes have equal loss. 
In the more familiar case of dispersive coupling, the pseudo-Hermiticity and the PT symmetries overlap. However, when the coupling is complex, we found an intriguing structure, where the reality of the spectrum is along hyperbolic lines in the regime diagrams. The relation between synchronization and pseudo-Hermiticity is not limited to just two lasers. This opens an arena for future study of complex band structures in laser lattices \cite{nixon2013observing}. In this case, the pseudo-Hermiticity symmetry can be seen as an imaginary flat-band analog \cite{leykam2018artificial}.

\section*{Materials and Methods}

\paragraph*{Experimental setup}
A schematic diagram of the experimental arrangement is presented in Fig. 2 (see Supplementary for a detailed version). It consists of two lenses that form a 4f telescope configuration, a gain medium, reflective spatial light modulator (SLM) and an output-coupler, with a total propagation length of 5m per roundtrip. The arrangement is essentially a perfect degenerate cavity laser \cite{cao2019complex}, where the field of each point at one end (the input) maps onto itself after a complete roundtrip. The gain medium is a doped Nd-YAG rod of 10-mm diameter and 11-cm length that is pumped by a xenon flesh lamp, generating a quasi-continuous-wave laser pulse of $200 \mu s $ duration. The reflective SLM \cite{house} at the input plane is used as a digital mask on which two holes of diameter $300 \mu m$ are imprinted. The field of each hole matches that of a single spatial Gaussian mode with waist of $ w_0 \approx 150 \mu m$ (see Supplementary), so the field of the holes can be viewed as that from two independent lasers. The SLM can thus be used to control the relative loss and frequency detuning of the two lasers. Coupling between the lasers is obtained by shifting the output-coupler from the exact imaging plane (more details below).

\paragraph*{Coupling mechanism}
The coupling between the lasers is generated by shifting the output-coupler by $\Delta z =10cm$ from the exact imaging plane. As a result, a portion of the light from each laser leaks to the other laser (see Fig. 2). The complex-valued coupling coefficients $\kappa_{nm}$ (by convention $\mathcal{H}_{nm}=-i \kappa_{nm}$) are given by the normalized overlap integral of the expanded field of one laser with that of the other one at the imaging plane (see Supplementary for explicit derivation).
As the field propagation from one laser to the other is symmetric, so are the coupling coefficients.
The effective Hamiltonian is therefore a complex-symmetric matrix, while generally not Hermitian, i.e.  $\mathcal{H}_{nm}=\mathcal{H}_{mn}$ but $\mathcal{H}^*_{nm} \neq \mathcal{H}_{mn}$.
In our experiments, we vary the coupling by changing the distance between the lasers (reflective spots on the SLM), so as to generate a dispersive, dissipative or complex coupling between the lasers.

\paragraph*{Detection arrangement}
One mirror serves as an output coupler where light emerges and propagates to the detection arrangement. The measurement of the lasers' amplitudes and relative phase is carried out by using an interferometer, schematically depicted in figure S1 of the Supplementary. In one arm of the interferometer, one laser is selected and expanded using a pinhole and a lens, to serve as a reference field. In the second arm, the laser field on the SLM is imaged by a 4f telescope. The light from both arms is then recombined on a CMOS detector with a small angle, resulting in interference fringes on top of each laser field. Fig. 1(b) shows a typical interference image. 
%The amplitudes of the lasers are obtained from the DC component while the phase is  of the image.
Each data point is averaged over 10 measurements.

\paragraph*{Laser rate equations}
Consider an array of many coupled lasers. The dynamics of the laser field and the gain medium is given by \cite{rogister2004power}:
\begin{eqnarray}
\frac{d E_m}{dt}   &=&   \frac{1}{\tau_p} \left[ (G_m-\alpha_m - i \Omega_m ) E_m - \sum_{n} \kappa_{mn} E_n \right]  \label{e1a} \\  
\frac{d G_m}{dt}   &=&   \frac{1} {\tau_c}\left[P_m-G_m \left(1+\frac{\left|E_m\right|^{2}}{I_{\mathrm{sat}}}\right)\right] \label{e1g}
\end{eqnarray}
where $\alpha_m,\Omega_m,P_m$ are the loss, frequency and pump of each laser, $\kappa_{nm}$ is the coupling matrix and $\tau_p,\tau_c,I_{\mathrm{sat}}$ are the cavity round trip time, the gain medium fluorescence lifetime and the saturation intensity.
We write the equations in a dimensionless form by rescaling the units of the electric field and time by setting $ I_{\mathrm{sat}} = 1 $ and $\tau_p =1$. 
For our symmetric coupling, we can write $\kappa_{12}=\kappa_{21} \equiv i \kappa  \mathrm{e}^{i \beta }$.
Using $\alpha_{1,2}= \alpha_0 \pm \Delta \alpha$ and $\Omega_{1,2}= \Omega_0 \pm \Delta \Omega$, \Eq{e1a} takes the form of \Eq{e6} for the two lasers, with $\mathcal{H}$ given by \Eq{e2}.
The initiation of lasing typically begins with a relaxation oscillations period \cite{siegman1986lasers}. When (and if) the system reaches a synchronized steady state, the gain \Eq{e1g} takes the value $G_m  = P_m/(1 + | E_i |^2)  $.
For a pump strength that is slightly above the lasing threshold value, the intensity is weak
and we can approximate the steady state gain  by $G_m \approx P_m$.
If we also assume uniform pumping $P_m=P$, the gain terms in \Eq{e2} are proportional to the unit matrix, and the fixed points can be approximated by the eigenstates of $\mathcal{H}$.
When both lasers have the same intensity, the eigenstates of $\mathcal{H}$ are exact solutions, since $G_1=G_2$, irrespective of the pump strength or the intensity.

\paragraph*{Anyonic PT symmetry condition.}
Here we show explicitly the condition for our effective Hamiltonian to have anyonic-PT symmetry.
In order to avoid confusion, here we replace $\beta$ by $\tilde{\beta}$ in the anyonic commutation \Eq{e1}.
We define the time-reversal and parity operators in the standard way \cite{nimrod1}, such that $\mathcal{T}$ perform complex conjugation and the parity operator here is given by
% $ \mathcal{P} = 
% \big(\begin{smallmatrix}
%   0 & 1\\
%   1 & 0
% \end{smallmatrix}\big)$.
% %
%
\begin{eqnarray} 
\mathcal{P} \ = \  
\left( 
\begin{array}{cccc}
 0   &  1 \\
 1 & 0    
\end{array}
\right) .
\end{eqnarray}
Additionally, we have $\mathcal{P}^2=1, \mathcal{T}^2 = 1$ and $[\mathcal{P},\mathcal{T}]=0$. 
We can rewrite \Eq{e1} as 
\begin{eqnarray} 
\mathcal{H} \ = \ \mathcal{PT} \mathrm{e}^{-2i \tilde{\beta} } \mathcal{H PT}  \ =  \  \mathrm{e}^{2i \tilde{\beta} }   \mathcal{PH^* P}  .
\end{eqnarray}
For the Hamiltonian of \Eq{e3}, we have
\begin{eqnarray}
\left( 
\begin{array}{cccc}
 z   & \kappa \mathrm{e}^{i \beta } \\
 \kappa \mathrm{e}^{i \beta } & -  z   \\
\end{array}
\right) 
\ = \ 
\mathrm{e}^{2i \tilde{\beta} }
\left( 
\begin{array}{cccc}
 -z^*   & \kappa \mathrm{e}^{-i \beta } \\
 \kappa \mathrm{e}^{-i \beta } &   z^*   \\
\end{array}
\right) .
\end{eqnarray} 
The off-diagonals give $ \mathrm{e}^{i \beta } =  \mathrm{e}^{i (2 \tilde{\beta} - \beta)}   $, which means we must have $ \tilde{\beta} = \beta $ (optionally $ \tilde{\beta}$ and $\beta$ can differ by $\pi$, but this makes no difference), as in our original definition of \Eq{e1}.
The diagonal terms give $ z= - z^* \mathrm{e}^{2i \beta}  $ or $ Re [z \mathrm{e}^{-i \tilde{\beta}}]=0$, so that with $z=\Delta \Omega - i \Delta \alpha$ we obtain the anyonic-PT symmetry line \Eq{e3}.

The anyonic-PT symmetry can also be understood as a generalization of the more familiar PT-symmetric case. Every anyonic-PT symmetric Hamiltonian can be written as $ \mathcal{H}= \mathrm{e}^{ i \beta} \mathcal{H}_{PT}  $, where $ \mathcal{H} _{PT}  $ is PT symmetric. This is because \Eq{e1} is formally equivalent to $[ \mathcal{PT}, \mathrm{e}^{ -i \beta} \mathcal{H}]=0$.
The celebrated PT-symmetric 2-site system (see e.g. \cite{bender2007making}), is given in our notation by
\begin{eqnarray}
\mathcal{H}_{PT} \ = \ 
\left( 
\begin{array}{cccc}
 \Delta \alpha    & \kappa \\
 \kappa & -  \Delta \alpha   \\
\end{array}
\right) 
\end{eqnarray}
which is a special case of \Eq{e2} with purely real coupling ($\beta=0$) and vanishing frequency detuning ($\Delta \Omega=0$).
When multiplying by the phase factor $\mathrm{e}^{ i \beta} $, the coupling becomes complex as required, while the line $\Delta \Omega=0$ is rotated in the $(\Delta \alpha , \Delta \Omega)$ plane.

\paragraph*{Synchronization borders.}
To derive the synchronization regimes of Fig. 4, we first re-write \Eq{e1a} in terms of the amplitudes and phases using $E_i=A_i \mathrm{e}^{i \phi_i}$, to obtain \Eq{e7} for the dynamic of the phase difference $\phi$.
Frequency synchronization essentially means the lasers are phase-locked, i.e. $d\phi/ dt=0$.
The steady-state $\phi$ depends on the amplitude ratio, $\beta$, $\kappa$ and $\Delta \Omega$, within some allowed frequency range. To find the border $\Delta \Omega$ values for which synchronization is possible, we maximize the last two terms of \Eq{e7} with respect to $\phi$.
Taking the $\phi $ derivative gives $A_1^2 \sin (\phi + \beta)=A_2^2 \sin (\phi - \beta)$, which solves for the relative phase at the synchronization border. Using this condition and \Eq{e7}, we find the maximal $(\Delta\Omega)^2$ is given by:
\begin{eqnarray}
\max_\phi \left[  \frac{A_1}{A_2} \cos(\phi + \beta ) - \frac{A_2}{A_1} \cos(\phi - \beta ) \right]^2 \nonumber \\
\ = \
\left( \frac{A_1}{A_2}  \right)^2 + \left( \frac{A_2}{A_1}  \right)^2 - 2 \cos 2 \beta .
\end{eqnarray}
Hence, synchronization is possible in the frequency detuning range \Eq{e9}. %\cite*{}

%\section*{References}

%%%%%

\section*{Acknowledgments}

We thank Stefan Rotter, Nimrod Moiseyev, Adi Pick, and Eilon Poem for useful discussions.

%%%%%

%\bibliographystyle{ScienceAdvances}

%\bibliography{PTbib}

%\begin{scilastnote}
%\item test
%\end{scilastnote}

\end{document}